\documentclass[english,prl,twocolumn,superscriptaddress,showpacs,preprintnumbers,amsmath,amssymb]{revtex4}
\usepackage[T1]{fontenc}
\usepackage[latin1]{inputenc}
\usepackage{graphicx}
\makeatletter
\usepackage{babel}
\makeatother

\begin{document}

\title{Criticality of Vacancy-Induced Metal-Insulator Transition in Graphene}

\author{Shangduan Wu}
\address{Department of Physics and Astronomy, University of Utah, Salt Lake City, Utah 84112, USA}

\author{Feng Liu}\thanks{E-mail: fliu@eng.utah.edu}
\address{Department of Materials Science and Engineering, University of Utah, Salt Lake City, Utah 84112, USA}

\date{\today}

\begin{abstract}

The criticality of vacancy-induced metal-insulator transition (MIT)
in graphene is investigated by Kubo-Greenwood formula with
tight-binding recursion method. The critical vacancy concentration
for the MIT is determined to be $x_{c} \approx 0.053\%$. The scaling
laws for transport properties near the critical point are examined
showing several unconventional 2D localization behaviors. Our
theoretical results have shed some new lights to the understanding
of recent experiments in H-dosed graphene [Phys. Rev. Lett.
\textbf{103}, 056404 (2009)] and of 2D disordered systems in
general.

\end{abstract}


\maketitle

Graphene, a single atomic layer of graphite, has drawn much interest
in the last few years \cite{Sci04-Nov}. Because of the linear
dispersion relation near the Dirac point, many unconventional
transport behaviors have been shown in the pristine 2D graphene
\cite{quantum_hall,Nat07-Nov,weak_localization}. On the other hand,
defects and impurities are often present in graphene, so their
effects on graphene transport properties are of significant
scientific interests and practical implications. Vacancy, one of the
natural defects present in graphene, as observed in experiment
\cite{Nat04-Has}, has been a subject of intense theoretical study
\cite{PRL06-Per,PRB06-Pere,PRB08-Per,PRB08-Huang,PRB08-Wu,PRB08-Pal,PRB08-Sch}.
Vacancy is predicted to induce \emph{quasi}-localized states in the
vicinity of Fermi energy \cite{PRL06-Per,PRB08-Per,PRB08-Wu},
leading to a metal-insulator transition (MIT). However, some
fundamental properties, especially the criticality of the
vacancy-induced MIT in graphene remains unclear. One important
question is what is the critical vacancy concentration for MIT.
Furthermore, what are the scaling laws characterizing the transport
properties close to the critical point? The answers to these
questions will help us not only to explain recent experiment results
in graphene \cite{PRL09-Bos} but also to better understand, in
general, localization and transport behaviors in 2D disordered
systems.

In this Letter, we report a theoretical study of criticality of the
vacancy-induced MIT in graphene. Most importantly, we determine the
critical vacancy concentration for MIT to be $x_{c}\sim 0.053\%$,
which may be compared to the experimentally observed critical H
coverage of $\sim 0.08 \%$ for the H-dosing-induced MIT in graphene
\cite{PRL09-Bos}. We found that electrical conductivity
$\sigma_{xx}$ scales with the diffusion length $L$ in a power-law,
in contrast to the logarithmic scaling predicted for a conventional
2D localization behavior \cite{PR58-And, RMP85-Lee}. Above $x_c$,
there exists a mobility edge $E_c$ defining a transition point of
energy from the localized states to the extended states, and the
transition is discontinuous as characterized by a \emph{non-zero}
minimum conductivity $\sigma_{min}$ at $E_{c}$. $\sigma_{min}$ is
found to vary from $1$ to $1.3e^{2}/h$, different from the universal
value of $0.628e^{2}/h$ expected in a conventional 2D localization
system \cite{RMP85-Lee}. Our results are consistent with the most
salient features of the H-dosing-induced MIT in graphene observed
recently \cite{PRL09-Bos}, suggesting that both vacancy and H belong
to short-range disorder in graphene.

The graphene is described by the $\pi$-band, nearest-neighbor
tight-binding Hamiltonian as

\begin{eqnarray}
\hat{H}=\sum_{i,\alpha}\varepsilon_{i}|i\alpha\rangle\langle
i\alpha|-\sum_{\langle i,
j\rangle,\alpha}\gamma_{ij}|i\alpha\rangle\langle j\alpha|,
\end{eqnarray}

\noindent where \emph{i} and \emph{j} are the neighboring sites on
the lattice, $\alpha=(\uparrow,\downarrow)$ is spin index.
$\varepsilon_{i}$ is the on-site energy set to zero, and
$\gamma_{ij}=\gamma=2.7$ eV is the hopping energy. To calculate the
conductivity, the real-space recursion method \cite{recursion} in
Kubo-Greenwood formula is adopted, which has been proved to be a
powerful tool in treating various disordered graphene systems
\cite{PRL08-Roc1, PRL08-Roc2}. The DC conductivity $\sigma_{xx}$ at
zero temperature can be calculated as
$\sigma_{xx}(E)=(e^{2}/S)\rho(E)\lim_{t\rightarrow+\infty}\mathcal{D}(E,t)$.
Here $S$ is area per atom. $\rho(E)=Tr[\delta(E-\hat{H})]$ is
average density of states (ADOS). $\mathcal{D}(E,t)$ is diffusion
coefficient, which is defined as
$\mathcal{D}(E,t)=\frac{1}{t}\frac{Tr\{[\hat{x},\hat{U}(t)]^{+}\delta(E-\hat{H})[\hat{x},\hat{U}(t)]\}}{Tr[\delta(E-\hat{H})]}$,
where $\hat{x}$ is the \emph{x}-component of the position operator,
$\hat{U}(t)=\exp(-i\frac{\hat{H}t}{\hbar})$ is time evolution
operator, and $t$ is diffusion time. Periodic boundary conditions
are used with a supercell sizes $L_{x}\approx738$ nm and
$L_{y}\approx213$ nm. Vacancies are introduced simply by removing
the atoms from the lattice \cite{PRL06-Per, PRB08-Per}.

\begin{figure}[center]
\begin{center}
\includegraphics[width=7cm]{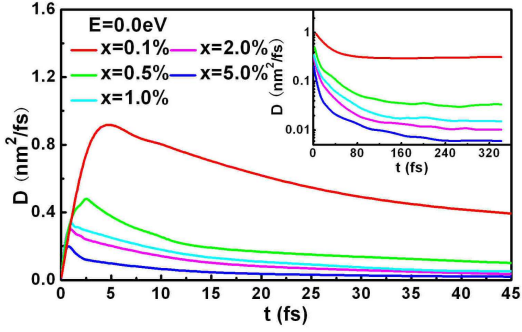}
\end{center}
\caption{(Color online) Diffusion coefficient ($\mathcal{D}$) as a
function of time ($t$) for different vacancy concentration ($x$) at
$E_F$.}
\end{figure}

Figure 1 shows the time dependence of diffusion coefficient
$\mathcal{D}$ at Fermi energy ($E_F=0.0$ eV) for different vacancy
concentration ($x$) in the range of $x\geq0.1\%$. The inset shows
the behavior of $\mathcal{D}$ at large time scale. Three main
features are observed: (1) $\mathcal{D}$ first increases linearly
with time and reaches its maximum $\mathcal{D}_{max}$ at a very
short time, less than $5$ fs. $\mathcal{D}_{max}$ is much smaller
than $\mathcal{D}=v_{F}^{2}t$ for perfect graphene, where Fermi
velocity $v_{F}=3ta/2\hbar\approx c/300$. This means that the
transport leaves the ballistic regime ($d\mathcal{D}/dt>0$) quickly
and enters the diffusive regime ($d\mathcal{D}/dt=0$) due to
vacancies. However, the diffusive regime is very short, without a
plateau often seen in other systems. (2) Beyond $\mathcal{D}_{max}$,
the system enters the localization regime, where $\mathcal{D}$
decays in a power law $\mathcal{D}\propto t^{-\lambda}$, similar to
the behavior of disordered carbon nanotude \cite{PRB04-Tri}. This
indicates that the system stays in the localization regime
($d\mathcal{D}/dt<0$) when $x\geq0.1\%$ at large time scale, in
agreement with the previous analytical results \cite{PRL06-Per,
PRB08-Per, PRB08-Wu}. (3) $\mathcal{D}$ decreases with the
increasing $x$ (inset of Fig. 1). This is because higher vacancy
concentration leads to stronger scattering, reducing the diffusion
coefficient. We note that at a given vacancy concentration, $\rho$
is a constant independent of time. Therefore, the conductivity
$\sigma_{xx}\sim\mathcal{D}\rho$ will have the same power-law
scaling with time, $\sigma_{xx}\propto t^{-\lambda}$, as
$\mathcal{D}$. Moreover, higher concentration will cause faster
decay (larger $\lambda$) in both diffusion coefficient and
conductivity.

\begin{figure}[center]
\begin{center}
\includegraphics[width=7cm]{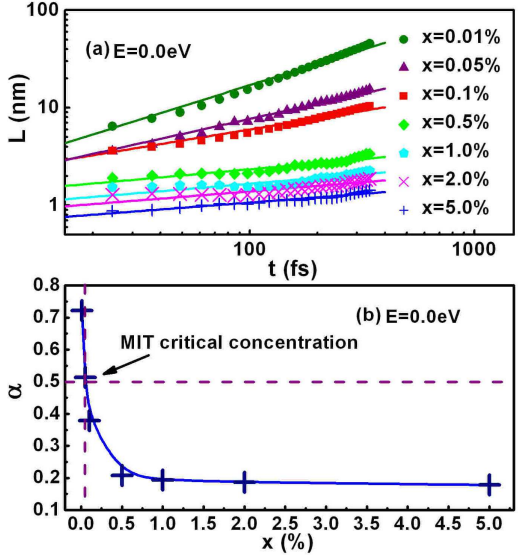}
\end{center}
\caption{(Color online) (a) Diffusion length ($L$) as a function of
time ($t$) for different vacancy concentration ($x$) at $E_F$. (b)
Scaling exponent $\alpha$ as a function of vacancy concentration
($x$) at $E_F$.}
\end{figure}

The relationship between conductivity and diffusion length is very
useful to better understand the transport behavior. By definition,
$L(E,t)=\sqrt{\mathcal{D}(E,t)t}$, then we can easily find that
$L\propto t^{\frac{1-\lambda}{2}}=t^{\alpha}$, which is verified for
all the concentrations simulated, as shown in Fig. 2(a) with a
Log-Log plot of diffusion length ($L$) versus time ($t$).
Substituting $t$ with $L$, $\sigma_{xx}$ scales with $L$ as
$\sigma_{xx}\propto L^{\frac{2\alpha-1}{\alpha}}$. By dimensional
analysis, $\sigma_{xx}$ can be expressed as

\begin{eqnarray}
\sigma_{xx}(L)=\sigma_{0} (e^{2}/h)(L/l)^{\frac{2\alpha-1}{\alpha}},
\end{eqnarray}
where $\sigma_{0}$ is the conductivity in the diffusive regime in
unit of $e^{2}/\hbar$ and $l$ is the electron mean free path. It is
important to note that such power-law dependence of conductivity on
diffusion length is different from the usual logarithmic dependence,
$\sigma(L)=\sigma_{0}-(e^{2}/\hbar\pi^{2})ln(L/l)$, as predicted by
2D scaling theory of localization \cite{RMP85-Lee}. We attribute
this difference to the distinct nature of the
\emph{quasi}-localization induced by vacancies. The wave-function
amplitude of the localized state induced by vacancies decays with
the distance as $1/ r$ \cite{PRL06-Per, PRB08-Per}, which is not
normalizable in 2D. Therefore, such \emph{quasi}-localized state
\cite{PRL06-Per, PRB08-Per} is different from the usual localized
state, whose wave function decays exponentially with the distance
\cite{RMP85-Lee}.

It has been predicted that the vacancy-induced
\emph{quasi}-localization will lead to MIT in graphene
\cite{PRL06-Per,PRB08-Per}. However, the critical vacancy
concentration where the MIT occurs remains unknown. Here, we will
determine the critical vacancy concentration for the MIT. From Fig.
2(a), we derive the scaling exponent, $\alpha$, from the linear
fitting of $L$ versus $t$, and the resulting $\alpha$ are plotted as
a function of $x$ in Fig. 2(b). The curve of $\alpha(x)$ can be fit
nicely by $L\propto t^{\frac{1-\lambda}{2}}=t^{\alpha}$, in the
whole range of concentration. Notice that when $\alpha=0.5$,
$\lambda=0$, then $d\mathcal{D}/dt=0$ because $\mathcal{D}\propto
t^{-\lambda}$, which means $\alpha=0.5$ defines the position of
diffusive regime. $\alpha>0.5$ ($\lambda<0$) means the ballistic
regime and $\alpha<0.5$ ($\lambda>0$) means the localization regime.
Thus, the power-law scaling holds true for all three transport
regimes. $\alpha$ initially decreases rapidly with the increasing
$x$, indicating the system quickly leaves the ballistic regime
entering the localization regime. The transition point at $\alpha
=0.5$ defines the critical vacancy concentration for MIT, which is
found to be $x_{c}\approx0.053\%$ in Fig. 2(b). This value seems to
agree quite well with the recent experimental result of critical
concentration of MIT ($0.08\%$) in H-dosed graphene
\cite{PRL09-Bos}. It implies that similar transport behavior exists
in these two disordered systems that both vacancy defect and H
impurity create similar localization phenomena in graphene. In
addition, the experiment suggested other unknown defects rather than
H may induce MIT at even lower defect concentrations
\cite{PRL09-Bos}, which might be attributed, at least partly, to
vacancies according to our calculation.

\begin{figure}[center]
\begin{center}
\includegraphics[width=7cm]{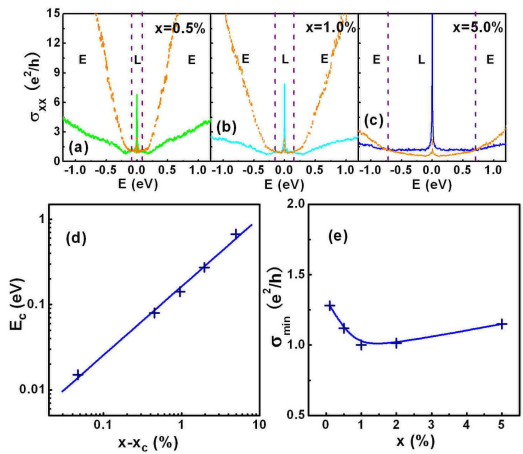}
\end{center}
\caption{(Color online) (a), (b) and (c) Conductivity
($\sigma_{xx}$) as a function of energy ($E$) for vacancy
concentration $x=0.5\%$, $x=1.0\%$ and $x=5.0\%$ at the time $t_1$
when $\mathcal{D}$ reaches its maximum (blue solid line) and
$t_2\approx 36.7$ fs (orange dash line). $``E"$ indicates the
extended state and $``L"$ indicates the localized state. (d) The
mobility edge ($E_{c}$) as a function of $x-x_{c}$. (e) The minimal
conductivity ($\sigma_{min}$) at $E_c$ as a function of $x$.}
\end{figure}

Next, we examine another criticality of the vacancy-induced MIT in
graphene, the mobility edge ($E_{c}$), which defines the critical
energy where the system changes from the localized state to the
extended state above the critical vacancy concentration $x_c$. We
only need to consider $x>x_{c}$, because when $x<x_{c}$ the
electronic states become extended over the whole energy range.
Figures 3(a), (b) and (c) show the energy dependence of conductivity
at two different times for three vacancy concentrations of $0.5\%$,
$1.0\%$, and $5.0\%$, respectively. One time is chosen at where
$\mathcal{D}$ reaches its maximum, $t_1(D=D_{max}$), and the other
is chosen at a much later time ($t_2\approx 36.7$ fs $>>t_1$). At
the low energies closed to $E_F$,
$\sigma_{xx}(t_1)>\sigma_{xx}(t_2)$, but at the high energies away
from $E_F$, $\sigma_{xx}(t_1)<\sigma_{xx}(t_2)$. This is because the
low-energy electronic states are localized whose conductivity
decreases with time; while the high-energy electronic states are
extended whose conductivity increases with time. This is in
qualitative agreement with the previous analysis of the inverse
participation ratio (IPR) of electron wavefunction \cite{PRL06-Per,
PRB08-Per}. Then, the mobility edge $E_{c}$ can be obtained as the
energy for $\sigma_{xx}(t_1)=\sigma_{xx}(t_2)$. In Figs. 3(a), (b)
and (c), we see that the mobility edge $E_c$ increases with the
increasing vacancy concentration, as the increasing vacancy
concentration expands the energy range of localization when
$x>x_{c}$.

In Fig. 3(d), we plot $E_{c}$ as a function of $x$, in a Log-Log
scale, which clearly shows a power-law dependence of $E_{c}$ on $x$.
Fitting the simulation data, we obtain the scaling relation of
$E_{c}\propto (x-x_{c})^{\beta}$, with the scaling exponent
$\beta\approx0.81$. Importantly, a \emph{non-zero} minimum
conductivity $\sigma_{min}$ appears at the mobility edge. Figure
3(e) shows that $\sigma_{min}$ first decreases and then increases
with $x$, but the reason for such trend is unclear. The existence of
$\sigma_{min}$ is possibly due to the \emph{quasi}-localized nature
of electronic states induced by vacancy in graphene. It indicates
that the transition from the localized states to the extended states
at the mobility edge is discontinuous as Mott argued
\cite{RMP85-Lee}, suggesting the break-down of one-parameter scaling
theory. Also, our simulated $\sigma_{min}$ varies from $1$ to $1.3
e^2/h$, different from the universal value of $0.628 e^{2}/h$ as
expected in conventional 2D localization system \cite{RMP85-Lee}.

\begin{figure}[center]
\begin{center}
\includegraphics[width=7cm]{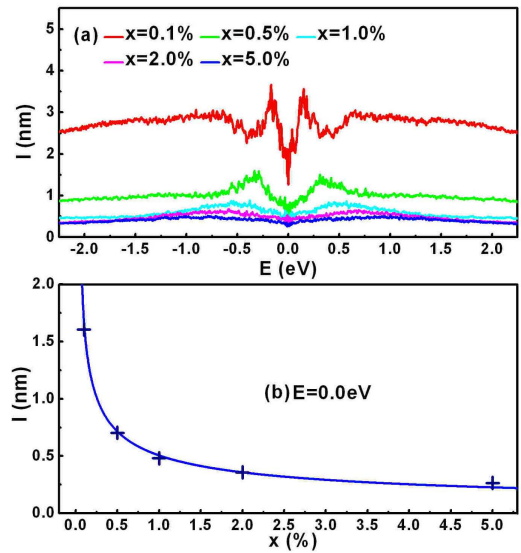}
\end{center}
\caption{(Color online) (a) Mean free path ($l$) as a function of
energy ($E$) for different vacancy concentration ($x$). (b) Mean
free path ($l$) as a function of vacancy concentration ($x$) at
$E_F$. The blue solid line is a fit to the data by $l\propto
x^{-1/2}$.}
\end{figure}

Last, we examine the electron mean free path $l$, an important
length scale characterizing the degree of electron scattering by
disorder. In Fig. 4(a), the energy dependence of $l$ is shown, which
is obtained by $l(E)=\mathcal{D}_{max}(E)/v(E)$ \cite{PRB04-Tri,
PRL97-Roc}.  $v(E)$ is the average wave-packet velocity at energy
$E$, extracted from the diffusion coefficient in the ballistic
regime at short times as $\mathcal{D}(E,t)\approx v(E)^{2}t$. It is
interesting to observe a minimal mean free path at $E_F$. This
distinguishes vacancies from other disorders in graphene that give
rise to a maximal mean free path at $E_F$ \cite{PRL08-Roc1,
PRL08-Roc2}. The minimal mean free path means that the lifetime of
electrons is the shortest near $E_F$ in the localization regime,
violating graphene's ordinary Fermi liquid behavior
\cite{Nat07-Bos}. Such an unusual behavior has also been
experimentally observed in graphene dosed with a high concentration
of H atoms \cite{PRL09-Bos}.

Near $E_F$, the mean free path $l$ is much less than the average
distance between vacancies ($d$). This is because the
\emph{quasi}-localized states substantially reduce $l$ in the
low-energy region. For example, when $x=0.1\%$, $l$ is found to be
$1.6$ nm, while $d$ is in the order of $5$ nm. In Fig. 4(b), we plot
$l$ as a function of $x$, which shows that $l$ scales with $x$ in a
power law $l=A x^{\delta}$ ($\delta=-0.5$) at $E_F$. Fitting the
simulation data, we obtained the parameter $A=0.508$ nm. This
power-law scaling has been derived before from the full
self-consistent Born approximation (FSBA) \cite{PRB06-Pere}, but the
FSBA is limited to extremely low concentration and can not describe
the localization behavior near $E_F$ and give the quantitative
values of $l$.

Our quantitative results indicate that vacancies are short-range
disorders similar to H atoms in graphene as suggested by a recent
experiment \cite{PRL09-Bos}. The similar results between vacancy
defect and H impurity can be understood by the dramatic
reconstruction of the energy spectrum near the Dirac point in both
cases, while the Fermi energy is preserved. As short-range local
scatters, both vacancy and H lead to strong localization by
significant scattering. We note that at low defect/impurity
concentration, the mean free path $l$ is shorter for vacancy ($1.6$
nm at $x=0.1\%$) than for H atom ($\sim4-5$ nm at $x=0.1\%$). This
is because vacancy produces a much stronger local scattering center
(creating an infinite barrier) than H (inducing a structural
transition from $sp^{2}$ to $sp^{3}$ hybridization). However, we
find that the rate of $l$ decreases faster with $x$ for H
($\delta<-1$) \cite{PRL09-Bos} than that for vacancy
($\delta=-0.5$). This is possibly because the scattering center
produced by vacancy is more localized limited to single atomic site,
while the scattering center produced by H atom is relatively more
extended as the $sp^{3}$ hybridization induces reconstruction in the
neighboring atoms around it. Consequently, at high concentration,
multiple scattering by H atoms becomes more and more important,
leading to a faster decay of $l$.  A different case was shown in
graphene substituted with boron or nitrogen, where the long-range
scattering is induced by chemical disorder with $\delta=-1$
\cite{PRL08-Roc2}.

In conclusion, the criticality of vacancy-induced MIT in graphene
has been investigated with quantitative simulations. The MIT is
shown to occur at the critical vacancy concentration
$x_{c}\approx0.053\%$, caused by the vacancy-induced
\emph{quasi}-localization. Several unconventional critical behaviors
have been revealed in the defected graphene with vacancies,
including the power-law rather than logarithmic scaling relation
between conductivity and diffusion length and the existence of a
minimum conductivity at the mobility edge that differs from the
universal value of $0.628e^{2}/h$ expected in conventional 2D
localization systems. Our results for vacancies are consistent with
the most salient features of the H-dosing-induced MIT in graphene as
observed in recent experiments \cite{PRL09-Bos}, suggesting that
both vacancy and H belong to the same class of short-range disorders
in graphene with H dosing involving stronger multiple scattering.
Our studies also shed some new lights to the general understanding
of critical behaviors of 2D disordered systems.

\end{document}